\documentclass[prd,aps,twocolumn,amsmath,amssymb]{revtex4}
\usepackage{graphicx}
\usepackage[T1]{fontenc}
\usepackage[latin1]{inputenc}
\usepackage{mathptmx}% instead of package times
\usepackage[scaled=0.9]{helvet} % or [scaled=0.92], if you like
\usepackage{courier}
\usepackage{amssymb}
\begin{document}
\title{Tomographic approach to resolving the distribution of LISA Galactic binaries} 
\author{Soumya D.~Mohanty}
\email{mohanty@phys.utb.edu}
\author{Rajesh K. Nayak}
\email{rajesh@phys.utb.edu}
\affiliation{Center for Gravitational Wave Astronomy, The University of Texas at Brownsville, 80 Fort Brown, Brownsville, TX 78520, U.S.A.} 
\begin{abstract}
The space based gravitational wave detector LISA
is expected to observe a large population of Galactic white dwarf
binaries whose collective signal is likely to dominate instrumental noise at observational frequencies in the range $10^{-4}$ to $10^{-3}$~Hz. 
The motion of LISA modulates the signal of each binary in both frequency and 
amplitude, the exact modulation depending on the source direction and frequency. Starting with the observed response of one LISA interferometer
and assuming only doppler modulation due to the orbital motion of LISA, we show how the distribution of the {\em entire} binary population
in frequency and sky position can be reconstructed  using a tomographic approach. 
The method is linear and the reconstruction of
 a delta function distribution, corresponding to an isolated binary, yields a point spread function (psf). An arbitrary distribution and its reconstruction
are related via smoothing with this psf. 
%The angular FWHM of the psf ranges from *  to *~steradians for binary frequencies in the range $10^{-4}$ to $10^{-3}$~Hz and a 1 year observation
%period. The FWHM of the psf
%in binary frequency is  of the order of $1/T$ where $T$ is the total observation period. 
Exploratory results are reported demonstrating the recovery of binary sources,
 in the presence of white Gaussian noise. 
\end{abstract}
\maketitle

\section{Introduction}
\label{introduction}

The Laser Interferometric Space Antenna (LISA)~\cite{LISA} is a space based 
interferometric gravitational wave (GW) detector, currently in design, with a launch date 
sometime around 2015. LISA will consist of three proof masses in drag free
orbits around the Sun. Passing GW signals will be detected by monitoring fluctuations in 
the nominal distance of $5\times 10^6$~km between the proof masses 
 using laser interferometry.  LISA will be most sensitive to GW signals
in the $10^{-4}$ to $1$~Hz band and is expected to observe a 
wide variety of GW sources during its 3 to 5 year mission liftime.

Among the sources predicted for LISA will be close white dwarf
binaries within the Galaxy, estimated to number
 around $10^8$ or more  in the low frequency range ($10^{-4}$ to $10^{-3}$ Hz) of 
the LISA sensitivity band~\cite{BH97}. 
LISA will be an all-sky monitor and the signals from long-lived GW sources from the entire sky will be 
summed in the response of the detector. The sum of the signals from all the
Galactic binaries, which we call the {\em binary population} signal, is expected to
dominate the instrumental noise of LISA at low frequencies and may act as a noise source from which
signals from other sources, including bright individual binaries, will need to be extracted.
On the other hand, the spatial and period distribution of LISA binaries
is in itself an important astrophysical observable for the 
LISA mission. Measuring the
 distribution function would yield information pertinent to studies of the Galactic structure, for instance.

The reconstruction of the distribution function of sources, given the superposed signal from all of them, is a 
critical data analysis problem for LISA. Intertwined with this task is that of removing as much of the binary 
population signal as possible from the data so that other types of sources in the same frequency range can be detected.
Note that the binary distribution function automatically
includes the case of an isolated binary, in which case it is a delta-function.
Hence, reconstruction of the distribution function can also identify well isolated individual binaries.
Several methods~\cite{CL2003,CR2005,RU2005} are currently being explored in the above context.

In this paper, we present the first steps of a promising approach to reconstructing the LISA binary distribution function.
 The approach, which we call {\em tomographic reconstruction}, is based on
the observation that, loosely speaking, the response of LISA to the binary population signal
is a Radon transform~\cite{RD1917} of the true binary distribution function. The origin of this lies in the doppler modulation of binary signals
due to the orbital motion of LISA. Given the measured 
response of LISA it is, therefore, possible to reconstruct the distribution
function by applying an inverse Radon transform to the response. 
%The Radon transform is at the core of tomographic
%imaging techniques routinely employed in fields ranging from medicine to seismology. It has also been applied in Astronomy
%to develop the technique of Doppler tomography. 

The main purpose of the paper is to demonstrate the connection between the response of LISA
and the Radon transform of the binary distribution function. 
We then explore how one of the most straightforward inversion methods for the Radon transform, namely {\em direct Fourier inversion},
can be used to  reconstruct
the binary distribution function.  
 Tomographic reconstruction is a {\em linear} operation on the data. Thus, it is fully characterized by the response
to a delta-function distribution or, in other words, by the {\em point spread function} (psf). The
reconstructed distribution is related to the true one via smoothing by this psf. 

We present exploratory results using a simple, and certainly naive, method to identify source candidates in
the reconstructed distribution function. We also use a
 much simplified scenario for demonstrating reconstruction. For instance, we only take the doppler modulation
 due to the orbital motion of LISA into account.
We find that, for an observation period of 1~year and frequencies near $2$~mHz, binaries having a
matched filtering signal to noise ratio (SNR) of $\sim 7$ or higher and a minimum spacing of 
about 10 frequency bins (1 bin $\simeq 3\times 10^{-8}$~Hz) can be distinguished fairly easily. 
This is in the range of the expected density of bright binaries~\cite{CR2005}. Binaries
can also be seen at separations of 4 to 6 frequency bins but the simple identification method
finds spurious sources arising from the superposition of the sidelobes of the psf. 
A more sophisticated identification method based on the use of deconvolution methods, such as CLEAN~\cite{hoegbom}
, is required in this case. However, we will address the latter issue in a future work
since it is a fairly non-trivial problem in itself. 
 
The rest of the paper is organized as follows. In Section~\ref{radontransform} we give an outline
of the Radon transform and the Direct Fourier Inversion method. 
Section~\ref{response} describes how the response to a distribution function of binaries can be 
expressed in terms of a Radon transform. The algorithm for tomographic reconstruction is presented
in Section~\ref{reconstruction}. Section~\ref{casestudies} contains exploratory results. 
\section{Radon Transform and its inversion}
\label{radontransform}

The Radon transform~\cite{RD1917} of a function $f({\bf x})$, ${\bf x}\in \mathbb{R}^n$, is defined in terms 
of the integrals of $f({\bf x})$  over
all possible hyperplanes of dimension $n-1$. 
The function $f({\bf x})$ can be reconstructed completely from its Radon transform provided 
integrals are available for all hyperplanes in the original domain. 

An important application area of Radon transform is in medical imaging where it is well known
as the CAT scan.
The absorption of X-rays
passing through the three
dimensional density distribution of biological matter is measured along many directions.
The absorption measurements are then inverted to yield the three dimensional distribution.
Tomographic methods where first applied in astronomy by Bracewell\cite{BW1956},
for imaging microwave emitting regions on the surface of the Sun. The Radon transform forms 
the basis of  the technique  of Doppler tomography~\cite{steeghs} which has been used to image accretion discs.

We will now describe those aspects of the Radon transform, and its inversion, that are of immediate use to us.
Readers can refer to an extensive collection of textbooks and monographs
on the subject such as~\cite{RNTRNS} for an in-depth treatment of the subject.

\subsection{Definition}

Let $\mathbf{x}=\left(x_{1},\, x_{2},\,\cdots x_{n}\right)$ be a
point in $\mathbb{R}^{n}$. Let $\widehat{ \xi}$ and $p$ define a
hyperplane containing ${\bf x}$,
\begin{equation}
p=\widehat{\bm \xi}\cdot\mathbf{x}=\xi_{1}x_{1}+\xi_{2}x_{2}+\cdots\xi_{n}x_{n}.\label{eq:pln}
\end{equation}
 The Radon transform, $\check{f}(p,\xi)$, of a function $f(\mathbf{x)}$ is
defined as
\begin{equation}
\check{f}(p,\widehat{\xi})=\int f(\mathbf{x})\delta(p-\widehat{\xi}\cdot\mathbf{x})d\mathbf{x}\label{eq:Radef}
\end{equation}
Thus $\check{f}(p,\widehat{\xi})$ is the integral of $f(\mathbf{x})$ over the hyperplane defined by $p$, $\widehat{\xi}$. This integral
is called a {\em projection} in the Radon transform literature.

There exist several inversion methods for reconstructing
$f(\mathbf{x})$ from its  radon transform
$\check{f}(p,\xi)$. (In fact, an inversion method was provided by Radon in the same paper where 
 the transform was first defined.)  Here, we
use the {\em direct Fourier inversion} method developed by Bracewell~\cite{BR1967,RAMLM71}  which is
based on a remarkable relation between the Fourier and  Radon transforms of a function.

\subsection{Relation to the Fourier Transform}

Let a point in Fourier
space be denoted by, $\mathbf{k}=(k_{1},\, k_{2},\,\cdots k_{n})$.
We adopt the following convention for the $n$-dimensional Fourier transform, $\mathcal{F}\!\!\! f$, of a function $f(\mathbf{x})$,
\begin{equation}
\mathcal{F}\!\!\! f (\mathbf{k})=
%\mathcal{F}_{n}f=
\int f(\mathbf{x})e^{-2\pi i\,\mathbf{k}
\cdot\mathbf{x}}\, d {\bf x}\label{eq:deff}
\end{equation}
Let $\mathcal{F}_1\!\!\check{f}(s,\widehat{\xi})$ be the one-dimensional Fourier transform,
\begin{equation}
\mathcal{F}_1\!\!\check{f}(s,\widehat{\xi}) := \int_{-\infty}^{\infty}\check{f}(p,\widehat{\xi})\, e^{-2\pi i\, sp}dp
\end{equation}
It can be proved that,
\begin{equation}
\mathcal{F}\!\!\!f(\mathbf{k}) = \mathcal{F}_1\!\!\check{f}(k,\widehat{k})\;.\label{eq:rel1}
\end{equation}
That is, the $n$-dimensional Fourier transform of a function can be obtained by a suitable one-dimensional
Fourier transform of its Radon transform.

\subsection{Direct Fourier Inversion Algorithm}
\label{sec:dfia}
\begin{figure}
\includegraphics[scale=0.55]{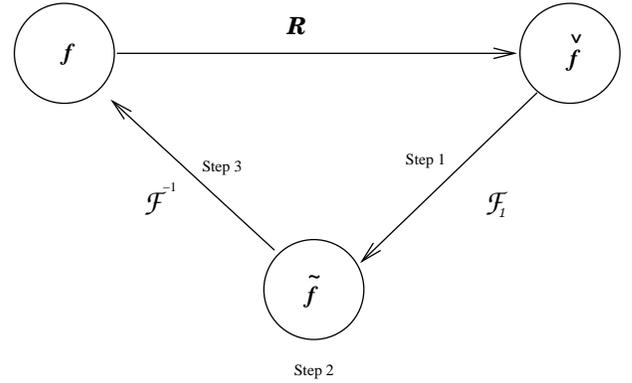}
\caption{The relation between the Radon and Fourier transforms of a function.\label{schematic_radon}}
\end{figure}

Eq.~\ref{eq:rel1} is the basis of the direct Fourier inversion algorithm for reconstructing $f({\bf x})$
from its Radon transform $\check{f}(p,\widehat{\xi})$. The basic steps of this algorithm are as follows.
\begin{enumerate}
\item Compute the one dimensional Fourier transform 
	${\mathcal F}_1\!\!\check{f}(k,\widehat{k})$,
for a  fixed $\widehat{k}$.
\item Repeat the above step for all possible values of $\widehat{k}$ to obtain
the Fourier transform $\widetilde{f}({\bf k}) = \mathcal{F}\!\!\!f({\bf k})$ in the Fourier domain.
\item Compute the inverse Fourier transform, $\mathcal{F}^{-1} \widetilde{f}$
to get $f({\bf x})$.
\end{enumerate}
Figure~\ref{schematic_radon} shows a schematic of the direct Fourier inversion method. 
\section{LISA response to the binary population signal}
\label{response}
In this section, we set up much of our basic notation. We describe, first, the response of LISA to a single binary
and then use it to obtain the response to a binary population signal. We assume that a binary emits a  
monochromatic GW signal even though
real binaries may not have circular orbits and, in addition, will have some
 frequency evolution due to orbital decay by GW radiation. In some binaries, accretion from one
star to another may drive frequency evolution. The assumption of a monochromatic signal should, however, be a
fairly good approximation for most binaries at frequencies $\sim 10^{-3}$~Hz or lower.
In any case, it is the simplest model to use for a first study such as this.
At low frequencies ($\leq 5$~mHz), where
 the binary population signal is expected, it is sufficient to use the LISA detector response 
 given in~\cite{CUT98}.

\subsection{Signal from a single binary}
We choose a Sun centered polar coordinate system with the orbital plane of the LISA centroid corresponding to polar angle $\theta=\pi/2$ and 
the azimuthal angle $\phi=0$ corresponding to the position of LISA at the start, $t=0$, of observations. 
Let $\widehat{w}=\left(\begin{array}{ccc}
\sin\theta\cos\phi, & \sin\theta\sin\phi, & \cos\theta\end{array}\right)$ be the unit vector pointing in the direction of source with angular
position $(\theta,\phi)$.
 Define unit vectors $\widehat{\theta}=\partial\widehat{w}/\partial\theta$
and $\widehat{\phi}=(1/\sin\theta)\partial \widehat{w}/\partial\phi$
transverse to $ \widehat{w}$.  The GW wave propagates in the
$-\widehat{w}$ direction and the metric perturbation is given by,
\begin{eqnarray}
h_{\mu\nu}(t) & = & h_{+}(t)\,\epsilon_{+,\mu\nu}+h_{\times}(t)\,\epsilon_{\times,\mu\nu},\label{eq:hmnbss}
\\
\epsilon_{+} & = & \widehat{\theta}\otimes\widehat{\theta}-\widehat{\phi}\otimes\widehat{\phi}, \\
\epsilon_{\times} & = & \widehat{\theta}\otimes\widehat{\phi}+\widehat{\phi}\otimes\widehat{\theta}.
\end{eqnarray}
where, $h_{+}(t)$ and $h_{\times}(t)$ are the waveforms of the two independent polarizations of a GW in the Transverse-Traceless gauge.
\begin{figure}
\includegraphics[scale=0.5]{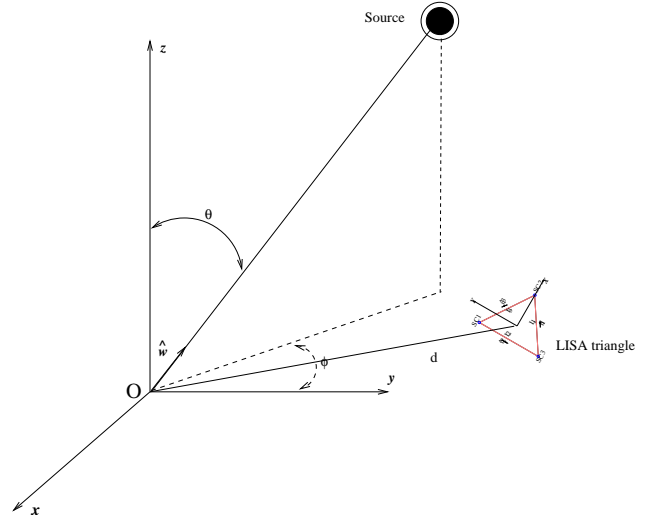}
\caption{\label{cap:Source-from-Barycentric}The coordinate 
convention used to describe a source in the barycentric frame.}
\end{figure}

For a monochromatic binary source, the polarization waveforms are given by,
\begin{eqnarray}
h_{+}(t) & = & A\left[A_{+}\cos2\psi\cos\Phi(t)+A_{\times}\sin2\psi\sin\Phi(t)\right],\label{eq:hp}\\
h_{\times}(t) & = & A\left[A_{\times}\cos2\psi\sin\Phi(t)-A_{+}\sin2\psi\cos\Phi(t)\right],\label{eq:hc}\\
A_{+} & = & \frac{1+\cos^{2}\epsilon}{2},\;\;A_{\times}  =  \cos\epsilon \;,
\end{eqnarray}
 where $\epsilon$ is the angle between $\widehat{w}$ and the 
orbital angular momentum vector of the binary, $\psi$ denotes the angle between $\widehat{\phi}$ and the
projection of the angular momentum vector on the plane transverse to $\widehat{w}$, and $A$ is an overall distance
dependent factor. The signal phase observed in the Sun centered frame is $\Phi(t) = \Omega t + \Phi_0$ where
$\Omega$ is twice the orbital frequency of the binary and $\Phi_0$ is the
initial phase at the beginning of the observation period.  

\subsection{Detector Response}

The orbits of the three LISA satellites are designed such that the
entire constellation will appear as a equilateral triangle in near rigid motion around the Sun. 
The plane of this triangle will be tilted at $60^\circ$ to the orbital plane of its centroid and the triangle will 
rotate once in its plane during a year. We neglect corrections,
 such as the ``flexing" of the whole configuration, to the basic picture above of LISA
orbital motion.

The signal phase observed at the LISA centroid will be doppler modulated due to orbital motion,
\begin{eqnarray}
\Phi (t) & = & \Omega t +\Omega \Phi_D(t) +\Phi_{0},\label{eq:phaselisa}\\
\Phi_D(t) & = & \hat{w}\cdot{\bf r}(t)\label{dopplerphase}\;,
\end{eqnarray}
where ${\bf r}$ is the position vector of the centroid of LISA, 
\begin{equation}
{\bf r}=R_{\odot}\left(
\cos\omega_{\odot}t, \sin\omega_{\odot}t,0
\right)
\;,
\end{equation}
 $R_{\odot}= 1\mathrm{AU}/c$ being the orbital radius of LISA, $c$ the speed of light, and $\omega_{\odot}=2\pi/T_{\odot}$,
the orbital period corresponding to $T_{\odot}=1$year. 

The detector response is given by the contraction of the signal tensor,
given by  Eq. (\ref{eq:hmnbss}), with the 
{\em detector tensor},
\begin{equation}
h_{IJ}(t)=h_{\mu\nu}:\mathcal{D}_{IJ}^{\mu\nu}\, .
\label{contract}
\end{equation}
where the detector tensor $\mathcal{D}_{IJ}$ is defined as follows.
 In the low frequency regime, where the armlength of the interferometer
are small compared to the wavelength of the GW signal, the
detector tensor can be written as the standard Michelson combination\cite{CUT98},
\begin{equation}
\mathcal{D}_{IJ}(t)=\widehat{n}_{I}(t)\otimes\widehat{n}_{I}(t)-\widehat{n}_{J}(t)\otimes\widehat{n}_{J}(t),
\label{dtensor_t}
\end{equation}
 where $\widehat{n}_{I}$ represents the unit vectors along the $I^{\mathrm{th}}$
arm of the interferometer. 
 In the following we consider only one out of the
two possible LISA interferometers. Hence, we drop the subscripts on $h_{IJ}(t)$ and $\mathcal{D}_{IJ}$.

\subsection{Response in terms of the Binary Distribution function}
\label{respdistfunc}
Let $\rho(\Theta)$ be the number density of the binaries as function
of parameters $\Theta= (
\theta, \phi, \Omega, \eta)$, where $\eta $ is the subset $(\epsilon,  \psi, A, \Phi_{0})$.
The number of binaries in the volume element $d\Theta$ at $\Theta$ is
given by 
$\rho(\Theta)d\Theta$.
We express $\rho(\Theta)$ as,
 \begin{equation}
\rho(\Theta) = \rho_\Omega(\Omega|\theta,\phi,\eta)
\rho^\prime(\theta,  \phi,\eta)\;,
\end{equation} 
where $\rho^\prime$ is $\rho$ marginalized over $\Omega$ and $\rho_\Omega$ is the 
conditional density of binaries in $\Omega$ for a given $(\theta,\phi,\eta)$.

For a given point $\theta$, $\phi$, $\eta$ in parameter space,
the   GW signal received at the origin of the Sun centered frame
due to binaries in the volume element $d\theta d\phi d\eta$ is
\begin{equation}
h_{+,\times}(t;\theta,\phi,\eta) =  \int_0^\infty d\Omega \rho_\Omega(\Omega|\theta,\phi,\eta) h_{+,\times}(t;\eta,\Omega)\;.
\end{equation}
where Eq.~\ref{eq:hp} and~\ref{eq:hc} provide the expressions for $h_+(t;\eta,\Omega)$ and $h_\times(t;\eta,\Omega)$
respectively.
We can decompose $h_{+,\times}(t;\theta,\phi,\eta)$ into a Fourier series,
\begin{equation}
h_{+,\times}(t_m;\theta,\phi,\eta) = \sum_{k=0}^{N-1} e^{i \omega_k t_m} s_{+,\times}[k;\theta,\phi,\eta]\;,
\end{equation}
where, for an observation period $T_{\rm obs}$, the frequency $\omega_k = 2\pi k/T_{\rm obs}$ and
$t_m = m T_{\rm obs}/N$, $N$ being the number of samples.

The response, $h(t;\theta,\phi,\eta)$,
of LISA to the GW signal coming from the volume element $d\theta d\phi d\eta$ can be written as (see Eq.~\ref{contract}),
\begin{widetext}
\begin{eqnarray}
h(t;\theta,\phi,\eta) &=& d\theta d\phi d\eta \rho^\prime (\theta,\phi, \eta) h_{\mu\nu}(t;\theta,\phi,\eta):\mathcal{D}^{\mu\nu}\;,\\
h_{\mu\nu}(t_m;\theta,\phi,\eta) & = & \sum_{a=+,
\times}
\epsilon_{a,\mu\nu}\left(\sum_{k=0}^{N-1} 
e^{i \omega_k (t_m+\Phi_D(t_m,\theta,\phi))} s_a[k;\theta,\phi,\eta]\right)\,,
\end{eqnarray}
\end{widetext}
where we have included the doppler modulation $\Phi_D$ due to the orbital motion of LISA
and $\mathcal{D}(t)$ is given by Eq.~\ref{dtensor_t}.
The total response of LISA to the binary population signal is,
\begin{eqnarray}
h(t_m) & =& \int d\theta d\phi d\eta\; h(t_m;\theta,\phi,\eta) \;\nonumber \\
& =& \int d\theta d\phi \!\!\sum_{a=+,\times}F_a(\theta,\phi,t_m) \times \nonumber\\
&&K_a\left(\theta,\phi,t_m+\Phi_D(t_m,\theta,\phi)\right)\;,\label{FK}\\
K_a(\theta,\phi,t_m) & = & \sum K_a [k;\theta,\phi] e^{i\omega_k t_m}\;,\label{dft_zdirection}\\
K_a[k;\theta,\phi] & = & \int d\eta \rho^\prime (\theta,\phi,\eta) s_a[k;\theta,\phi,\eta]\;,\label{marginalize_eta}\\
F_a(\theta,\phi,t_m) & = & \epsilon_{a,\mu\nu}(\theta,\phi):\mathcal{D}^{\mu\nu}(t)
\end{eqnarray}
Now, $F_a(\theta,\phi,t)$ is a completely known function, as is $\Phi_D(t,\theta,\phi)$, 
and, hence, it can be replaced by a function $G_a(\theta,\phi,t+\Phi_d(t,\theta,\phi)) = F_a(\theta,\phi,t)$. Thus, 
\begin{equation}
h(t_m) = \int d\theta d\phi\; K\left(\theta,\phi,t_m+\Phi_D(t_m,\theta,\phi)\right)\;,
\label{eq:refor1}
\end{equation}
where we have combined the integrand in Eq.~\ref{FK} into a single function $K(\theta,\phi,t)$. Since we do not 
know the binary distribution function, $K(\theta,\phi,t)$ is an unknown function.

\subsection{LISA Response as a Radon Transform}
\label{respradon}

It is clear from Eq.~\ref{eq:refor1} that the response $h(t)$ 
is obtained by integrating the function $K(\theta,\phi,t)$
 on the surface $(\theta,\phi,t+\Phi_{D}\left(\theta,\phi,t\right))$ parametrized by $\theta$ and $\phi$. 
Except for the fact that this surface is not a plane,  $h(t)$
is, therefore, nothing but the Radon transform (Sec.~\ref{radontransform}) of the three dimensional function $K(\theta,\phi,t)$.
To cast $h(t)$ as a standard Radon transform, we can make the following coordinate
transformation which turns the surface of integration  into a plane. Let ${\bf x}=(x,y,z)$,
\begin{equation}
\left(\begin{array}{c}x \\ y\\z\end{array}\right) = \left(\begin{array}{c}\sqrt{2} R_{\odot}\sin\theta\cos\phi \\
 \sqrt{2} R_{\odot}\sin\theta\sin\phi \\
\sqrt{2}(t+\Phi_D)
\end{array}\right).\label{eq:cartcord}
\end{equation}
In terms of the new coordinates and using Eq.~\ref{dopplerphase}, 
\begin{eqnarray}
t &=&\frac{z}{\sqrt{2}}-R_{\odot}\sin\theta\cos\phi\cos\omega_\odot t-\nonumber\\
&&R_{\odot}\sin\theta\sin\phi\sin\omega_\odot t,
\end{eqnarray}
and, hence, $
t=\xi\cdot\mathbf{x},
$
where
\begin{equation}
\xi=\frac{1}{\sqrt{2}}\left(\begin{array}{ccc}
-\cos\omega_\odot t, & -sin\omega_\odot t, & 1\end{array}\right),\label{eq:xit}
\end{equation}
In the new coordinates, 
\begin{equation}
h(t)=\int K(\mathbf{x})\delta(t-\xi\cdot\mathbf{x})d\mathbf{x}\;,
\label{resp_radon}
\end{equation}
which casts the LISA response in the form of a standard Radon transform. (The Jacobian associated with the coordinate transformation
has been absorbed in $K({\bf x})$.)

 Although the function $K(\mathbf{x})$, Eq.~\ref{resp_radon},  is not the 
same as the full distribution function $\rho(\Theta)$, it already informs us
 about how the binary distribution appears in the three most interesting parameters, namely, $\theta$, $\phi$ and
$\Omega$.  In fact, for the simple case of an isotropic detector antenna pattern the Discrete Fourier Transform of 
$K({\bf x})$ along the $z$ direction yields the distribution of binaries in frequency $\Omega$, for a given direction $\theta$, $\phi$,
marginalized over the subset of parameters $\eta$. 
(This can be seen
from Eq.~\ref{dft_zdirection} and~\ref{marginalize_eta} above.)
In the following we will restrict our attention to the reconstruction of $K(\mathbf{x})$ and by taking its
inverse Fourier transform along $z$, reconstruct the distribution of binaries along the spatial position $(x, y)$ and 
signal frequency $\Omega$.
\section{Tomographic reconstruction of LISA binary distribution}
\label{reconstruction}

Having shown that the response of LISA is a Radon transform of the binary distribution function (more precisely, its Fourier transform along
the source frequency), we look at
how and to what extent the latter can be reconstructed from the former.
 As mentioned earlier there are several inversion methods for the
Radon transform. In this paper we consider  the direct Fourier inversion method that 
was outlined in Sec.~\ref{radontransform}. 
%This inversion method may not be the optimal one
%for the particular problem at hand but it is a good starting point for further explorations.
 
We use a simplified scenario for demonstrating reconstruction:
The response of LISA to a binary signal is assumed to involve only the orbital doppler frequency modulation and,
 although LISA will have three interferometers, the data is assumed to be the output of only one interferometer. 
This  toy model is a conservative one since 
adding the amplitude and phase modulation
due to the rotating antenna pattern of LISA and bringing in a second interferometer can only bring improvements.

\subsection{Missing projections and approximate solution}
\label{sec:mpaas}

A Radon transform is invertible provided the integrals of the original function 
are known for {\em all} possible hyperplanes. The first problem we run into when inverting the LISA response is that
we have, in fact, very few projections available to us. This is simply because of the fact that we have only one orbit producing the
required doppler modulations for us. To construct more projections, we need multiple LISA detectors moving on different 
orbits around the Sun. 

The planes on which projections are available are specified by the unit normal $\xi$, defined in Eq.~(\ref{eq:xit}),
and the distance, $t$, of the plane from the origin. It is clear that $\xi$ is confined to a two dimensional cone and does not
cover all possible directions. 
Moreover, since $\xi$ is a function of $t$, we have only one plane
for any $t$ whereas the Fourier inversion method (Sec~\ref{radontransform})
 requires that for a fixed $\xi$, projections be available for
 all possible $t$. In the case of LISA, therefore, we have a severe
shortage of projections and correspondingly our ability to reconstruct source position and 
frequency will be limited.

It may appear that, given the above problem, the method of direct Fourier inversion is
inapplicable to the inversion of the LISA response. However, we can take advantage of the fact that 
$\xi(t)$  changes very slowly, compared to the period of the sources LISA is designed to detect,  
to make an approximation that allows direct Fourier inversion to be used: 
Assume that $\xi(t)$ is nearly constant over a sufficiently small time interval $\Delta$ ($\ll$ one year). 
The time series $h(t)$ (the LISA response) can be divided into segments of length $\Delta$
to each of which a constant $\xi$ can be associated, such as $\xi(t_0)$ where $t_0$ is the mid point of the segment.

Under the above approximation, for a given $\xi(t)$ we now have more than one
 projection corresponding to (nearly) parallel planes at distances $t \in [t_0-\Delta/2,t_0+\Delta/2]$
from the origin. (The sampling of the LISA response will give rise to a finite
number of projections.) Hence, we can now use the direct Fourier inversion method. 
The steps involved in the latter were
enumerated in Sec.~\ref{radontransform}. Translated to the particular case here, they are as follows.

\subsection{Numerical implementation of Direct Fourier Inversion}
\label{num_reconstruct}

First, the Discrete Fourier Transform of each segment (length $\Delta$) of $h(t)$ is computed. Each segment is obtained
  by multiplying $h(t)$ with 
a boxcar window function of length $\Delta$. 
(We ignore end effects caused by the smaller length of the last few segments.) Consecutive segments have an 
overlap of half the segment length.
The above step yields the 
the three dimensional Fourier transform, $\widetilde{K}({\bf k})$, of the function $K({\bf x})$ that is 
to be reconstructed (Fig.~\ref{schematic_radon}).

 The fact that $\xi$ is confined to a cone (Eq.~\ref{eq:xit}) restricts the Fourier
wave-vector ${\bf k}$ to a cone too. Define polar coordinates
$(k,\epsilon,\psi)$ in the Fourier space where $\epsilon$ is the 
polar angle measured from $k_z$ and $\psi$ is the azimuthal angle measured from $k_x$.  Then $\mathbf{k}$ is 
confined to the cone defined by
\begin{equation}
\mathbf{k}  = \frac{k}{\sqrt{2}}\left(\begin{array}{ccc}
-\cos \psi, & -\sin \psi, & 1\end{array}\right)\label{eq:cone}\;.
\end{equation}
The next step is to compute the three dimensional {\em inverse} Fourier transform of $\widetilde{K}({\bf k})$.
Transforming to polar coordinates, 
the expression for the
inverse Fourier transform can be written as, 
\begin{eqnarray}
	K({\bf x}) &= & \int \! \! \! dk\, \frac{e^{\sqrt{2}\pi i k z}}{\sqrt{2}} \! \! \int \!\! d\psi  \, k^2 e^{-\sqrt{2}\pi i k \left( x \cos\psi + y \sin \psi \right)}  
	\times\nonumber\\
&&  \widetilde{K}\left(k,
	\epsilon=\frac{\pi}{4}, \psi\right)\,. 
	\label{eq:distri}
\end{eqnarray}
Setting $\epsilon = \pi/4$ accounts for the restriction of the first fourier transform to the cone defined in Eq.~\ref{eq:cone}.
Having done the inverse Fourier transform above, we will obtain $K({\bf x})$. 

Recall that in order to obtain the distribution, $K(\theta,\phi,\Omega)$,
of the binaries  in frequency, $\Omega$, the Fourier transform of $K({\bf x})$ along $z$ needs to be computed for each
$(x,y)$ (which correspond to sky position). 
By definition, a doppler modulated monochromatic signal  with carrier frequency $f_0$ originating at $\theta$, $\phi$ 
corresponds to a pure sinusoidal function 
of $z$ with frequency $f_0/\sqrt{2}$ at the corresponding location $x$, $y$. Let the Fourier transform along $z$ of $K({\bf x})$ be
$g(k;x,y)$. Then $G(k;x,y) \propto g(k/\sqrt{2})$ is the desired source distribution along frequency,
\begin{eqnarray}
K({\bf x})  & = & \int \! dk \, e^{2\pi i k z} g(k;x,y) \propto \int \! dk \, e^{\sqrt{2}\pi i k z} G(k;x,y)\;,
\label{scaledft}
\end{eqnarray}
Thus, from Eq.~\ref{eq:distri} and~\ref{scaledft}, it follows that 
%\begin{widetext}
\begin{equation}
	K(x,y,k)=\int \!\! d\psi \,  k^2
	e^{-\sqrt{2}\pi i k \left( x \cos\psi + y \sin \psi \right)} \widetilde{K}\left(k,
	\frac{\pi}{4}, \psi\right)\,. 
	\label{eq:fxyk}
\end{equation}
%\end{widetext}
is the distribution function in sky position and frequency.
The integral over $\psi$ is approximated by a simple summation. (We have ignored overall
constant factors in the above derivation.)

\subsection{Point Spread Function of tomographic reconstruction}
\label{psf}

We have described the numerical implementation of the tomographic reconstruction method above. 
The method is explicitly linear, since it only involves two consecutive Fourier transforms. Consequently,
it can be fully characterized in terms of its response to a delta function distribution, i.e., the {\em point spread function} (psf).

The distribution function $K({\bf x})$ corresponding to a single monochromatic source 
is given by,
 \begin{equation}
K({\bf x})=\exp\left({ \frac{i \Omega_0 z}{\sqrt{2}}}\right)\delta(x-a)\delta(y-b)\;.\label{eq:dist}
\end{equation}
 Here, $(a,\, b)$ are the source location and $\Omega_0$ is the angular frequency of
the signal.
 The Radon transform $\check{K}(\widehat{\xi},t)$ of the above source distribution is,
\begin{equation}
\check{K}(\widehat{\xi},t) = \int d{\bf x} K({\bf x})\delta(t-{\bf x}\cdot \widehat{\xi})
= e^{i\Omega_{0}\left(t-a\xi_{1}-b\xi_{2}\right)}\label{eq:distRD}
\end{equation}
 where we have used the fact (c.f., Sec.~\ref{respradon}) that in the case of LISA, $\hat{\xi}$ is confined to a cone (c.f., Eq.~\ref{eq:xit}
).

 The first step in the inversion (c.f. Sec~\ref{sec:mpaas}) is 
to break the time series $h(t) = \check{K}(\widehat{\xi}(t),t)$ into sections of length $\Delta$. 
 The next step is to take the Fourier transform of each section. For the $i^{\rm th}$ section,
\begin{eqnarray}
\widetilde{K}({\bf k}_i) & = & \int_{t_{i}}^{t_{i}+\Delta}e^{i\Omega_{0}(t-a\xi_{1}-b\xi_{2})}\: e^{-2\pi ikt}dt\;,\nonumber\\
 & = &\Delta e^{-i\Omega_{0}(a\xi_{1}+b\xi_{2})} e^{2\pi i(f_{0}-k)(t_{i}+\Delta/2)}\times\nonumber\\
&& \mbox{sinc}\left(\left(f_{0}-k\right)\Delta\right)\label{eq:fftrd1}\;,\\
{\bf k}_i & = & k \widehat{\xi}_i \nonumber \;,
\end{eqnarray}
where $\widehat{\xi}_i = \widehat{\xi}(t_i + \Delta/2)$.
Finally, the three dimensional inverse Fourier transform is computed as described in Sec.~\ref{num_reconstruct}.
In polar coordinates,
\begin{eqnarray}
\widetilde{K}\left(k,\epsilon =\frac{\pi}{4},\psi\right)& =& \Delta e^{-i\sqrt{2}\pi f_{0}(a\cos\psi+b\sin\psi)}
\times\nonumber\\
&&\exp\left({2\pi i(f_{0}-k)\left(\frac{\psi}{\omega_\odot}+\frac{\Delta}{2}\right)}\right)
\times 
\nonumber\\
&& \mbox{sinc}\left(\left(f_{0}-k\right)\Delta\right)\;,
\label{eq:kpolarcoord}
\end{eqnarray}
where we have used the fact that $\psi = \omega_\odot t$, $\omega_\odot$ being the 
orbital period of LISA (c.f., Eq.~\ref{eq:xit}).
For source frequencies that are integral multiples of $2\pi/\omega_\odot$, 
substituting $K(k,\epsilon,\psi)$ obtained above into Eq.~\ref{eq:fxyk} gives
\begin{eqnarray}
K(x,y,k) &=&\frac{2\pi i^{n}\Delta k}{\sqrt{2}} \exp\left({2\pi i\left(f_{0}-k\right)\left(\frac{\Delta}{2}+\frac{\chi}{\omega_\odot}\right)}\right)\times \nonumber\\
&&{\rm sinc}\left(\pi\left(f_{0}-k\right)\Delta\right)J_{n}\left(2\pi\rho\right)\;,\label{eq:fdistri}\\
\rho\cos\chi & = & \frac{1}{\sqrt{2}}(k x - f_0 a)\;,\\
\rho\sin\chi &=&  \frac{1}{\sqrt{2}}(k y - f_0 b)\;,\\
n & = & \frac{2\pi (f_0 - k)}{\omega_\odot}\;,
\label{besselorder}
\end{eqnarray}
 where $J_n(x)$ is the Bessel function of the first kind of order $n$~\cite{AS72}. 
(There is no convenient representation for frequencies that are non-integral
multiples of $2\pi/\omega_\odot$.)
 
An important feature of the psf is the dependence (Eq.~\ref{besselorder}) of the order $n$ of  $J_n$ on 
frequency $k$. At the source frequency, the psf behaves like $J_0(2\pi r)$ in the $(x,y)$ plane,
where $r$ is the distance from the correct source location. However, when $k \neq f_0$, the behaviour
is that of $J_n(x)$, $n\neq 0$. Now, 
the first maximum of $J_n(x)$ occurs at $x\simeq n$. Therefore, at the correct source
location, the psf falls to zero rapidly as one moves away from the source frequency. This fall off is much faster
than the ${\rm sinc}\left(\pi\left(f_{0}-k\right)\Delta\right)$ dependence. 
%(Fig.~\ref{besseljn} shows $J_n(x)$ for $n=0,1,\ldots,10$.)
Thus, although naively one may expect a
frequency resolution no better than $\sim 1/\Delta$ due to the sectioning of the data, the frequency resolution
is actually $\sim 1/T_{\rm obs}$. 

\section{Exploratory results}
\label{casestudies}
The psf described above has sidelobes that spread a point source distribution upon reconstruction.
The sum of several sources, therefore, leads to maxima in the reconstructed density which
do not correspond to a real source. It should be possible to obtain a cleaner
reconstruction by applying sophisticated deconvolution methods such as CLEAN~\cite{hoegbom}.  
However, this is a fairly non-trivial task which needs to be addressed carefully. We leave this
to a future work and  report
only exploratory results in this paper. 

\subsection{LISA response and reconstruction parameters}
\label{results_params}
The time series $h(t)$ representing the response of LISA to the binary population signal is generated as follows.
Doppler modulated sinusoidal signals,
\begin{equation}
s_j(t) = A \cos\left(\Omega_j \left(t - x_j \xi_1(t) - y_j \xi_2(t)\right)\right)\;,
\end{equation}
are generated for a set of randomly chosen sky locations $(x,y)$ (c.f., Eq.~\ref{eq:cartcord}
) and frequencies $\Omega$. The spacing $\Delta k$ between consecutive values of $k = \Omega/2\pi$
values is chosen to be $b_{\rm min}\delta k \leq \Delta k  \leq b_{\rm max}\delta k$, where $\delta k = 1/T_{\rm obs}$, $T_{\rm obs}$
being the observation period (taken to be 1~year). (The examples shown here correspond to different 
choices for the integers $b_{\rm min}$ and $b_{\rm max}$.) We ignore the initial phase of the signal because we do not
pursue reconstruction of the distribution in parameters beyond $(x,y,\Omega)$. 
The binary population
signal is,
\begin{equation}
h(t) = \sum_j s_j(t)\;.
\end{equation}
We take $j=1,\ldots,100$ and $k_1=2$~mHz. 
The distribution function is then reconstructed only in a range $32\times \delta k$ centered at a frequency 
$\geq 10\times \delta k +k_1$. This procedure accounts for  the influence of binaries lying outside the frequency band
of immediate interest. 
All binary signals have the same matched filtering signal
to noise ratio (snr) $\rho$, defined for white noise as,
\begin{equation}
\rho = \sqrt{\sum_i s^2[i]}\;.
\end{equation}  
For the reconstruction, we split $h(t)$ into segments that are $\Delta = 1$~day long (c.f. Sec~\ref{reconstruction}) with
an overlap between consecutive segments of half this length. The sampling frequency
of the data is set at $0.1$~Hz. 

\subsection{Reconstruction examples}
In the following, we use $|K(x,y,k)|$ (Eq.~\ref{eq:fxyk}) to study tomographic reconstruction.
We use a simple method for identifying sources: First, $|K(x,y,k)|$ 
is obtained
 for LISA response consisting of only white, Gaussian noise.
 From this, we estimate a value $\eta$ such that it is the smallest supremum over all values of $ |K(x,y,k)|$ 
in a specified range. 
Each sky map $|K(x,y,k)|$, obtained for data containing both noise and the binary population
signal $h(t)$, is then normalized by $\eta$. (The same noise realization
is used in both the noise only and noise plus signal reconstructions.)
A threshold is then applied to the normalized sky maps and
any isolated cluster of pixels  in the $(x,y)$ plane with values greater than the threshold is taken to be an individual source.

\begin{figure*}
\includegraphics[scale=0.5]{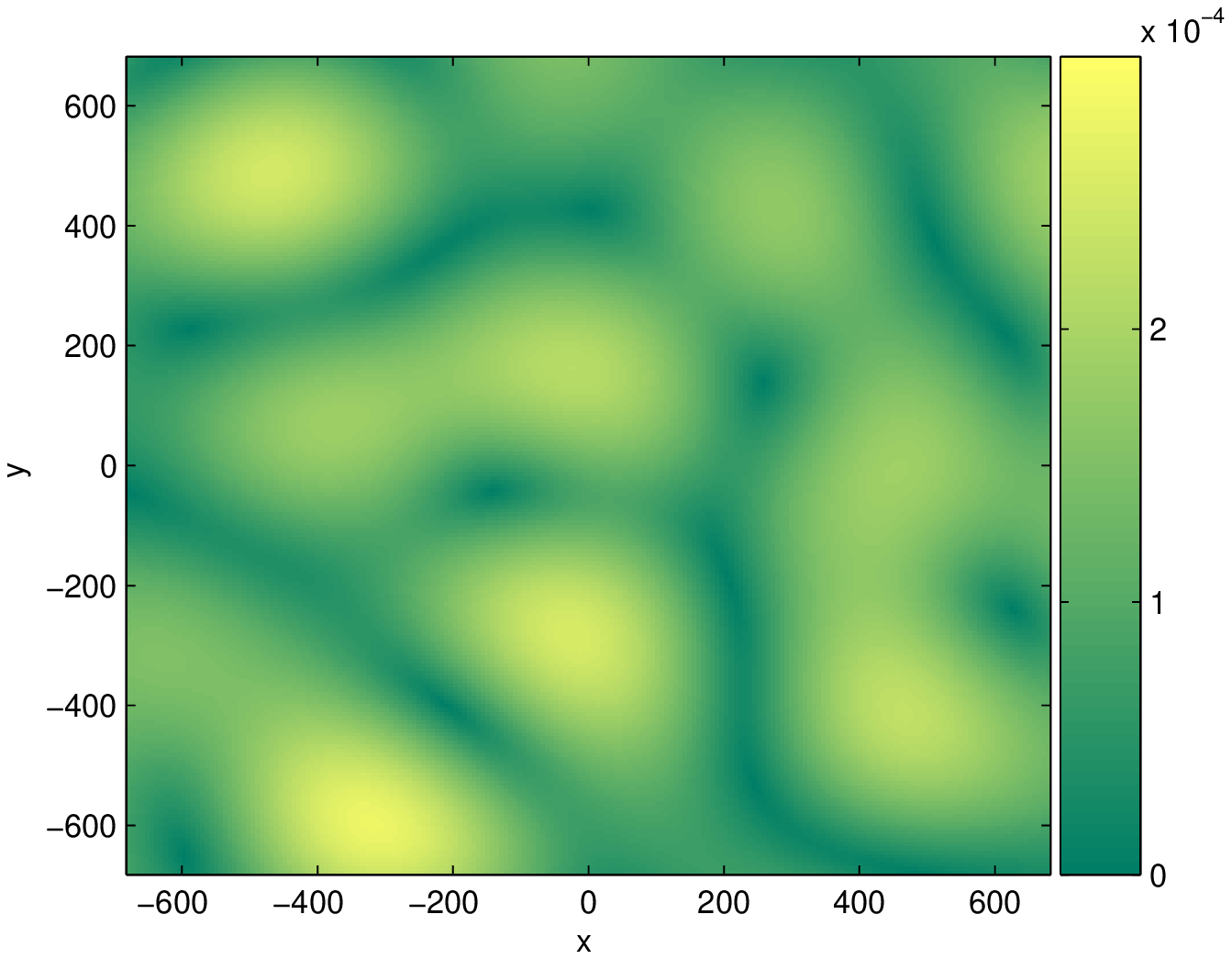}
\includegraphics[scale=0.5]{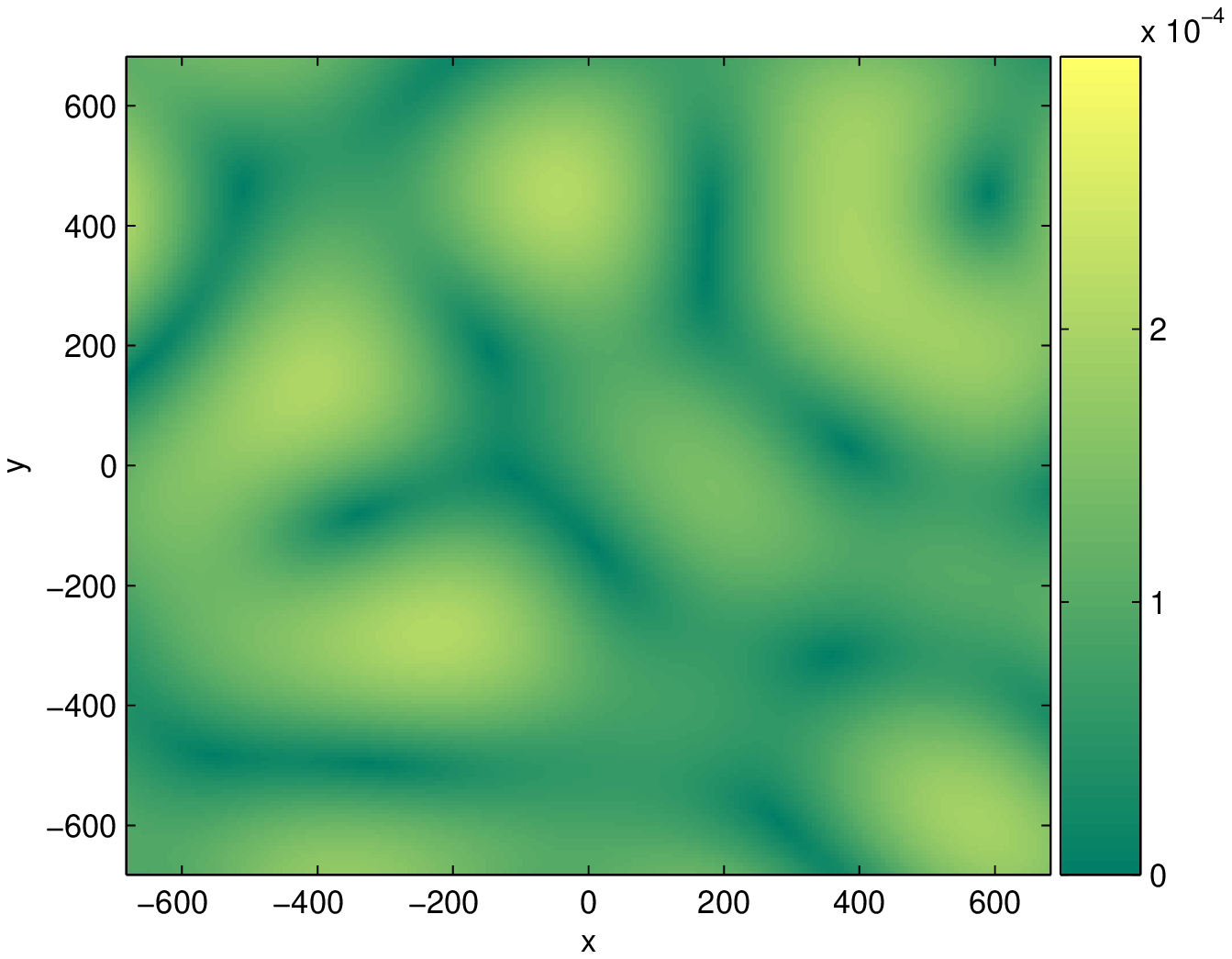}
\caption{Typical reconstructed sky maps for white, Gaussian noise. The two sky maps correspond to frequencies separated by
 $13\times \delta f$ (c.f., Sec~\ref{results_params}). \label{noise_reconstruction_singlepanel}}
\end{figure*}

\paragraph{Noise only --}
Fig.~\ref{noise_reconstruction_singlepanel} shows $|K(x,y,k)|$ at two widely separated frequencies $k_1$ and $k_2$, $k_2 - k_1 = 13\times \delta k$,
 when $h(t)$ is a realization of  white, Gaussian noise. 
From the values of $|K(x,y,k_i)|$ over the bandwidth $k_i \in [k_1+10\times \delta k, k_1+42\times \delta k]$ mentioned above, 
we estimate $\eta$, the smallest supremum of all pixel values in this frequency band  (and all values of $(x,y)$)
to be $\eta = 3\times 10^{-4}$. (The overall scale is arbitrary here.)
The effect of the psf is apparent from the presence of slow variations in the amplitude of the reconstructed distribution. 

\paragraph{Noise plus signal from well isolated, strong binaries --}
This example demonstrates the recovery of fairly large snr binaries that are well isolated in frequency. 
Fig.~\ref{snr15_10_to_15_1} and~\ref{snr15_10_to_15_2}
 show $|K(x,y,k)|/\eta$ for a realization of LISA response that has a binary population signal $h(t)$ added to 
 the same noise realization as Fig.~\ref{noise_reconstruction_singlepanel}. For the signal we choose $\rho=15$, $b_{\rm min}=10$
and $b_{\rm max}=15$. 

After normalization with $\eta$, a threshold of $|K(x,y,k)|/\eta =4$ is applied to all the sky maps. The contours
of $|K(x,y,k)|/\eta$ greater than this threshold are superimposed on the sky maps. We find that only the maximia
associate with the true source locations cross the threshold. (There are two minor peaks that cross the threshold
but are not associated with the main lobe of the psf.) The maxima of $|K(x,y,k)|$ do not coincide exactly with the 
true source location. This is due to the fact that the frequencies of the  sources are not exactly coincident with the
those associated with the maps.
(The binary frequencies need not correspond exactly to  integral multiples of $\delta k$. When this is the case, we show the source location in 
 the two sky maps on either side of the true source frequency.)

Another feature to note in the figures is the
 rapid falloff of the psf at the source location with a shift of one frequency bin from the source
frequency. This
behavior is expected from the analytical form given in Sec.~\ref{psf}.
\begin{figure*}
\includegraphics[scale=1.0]{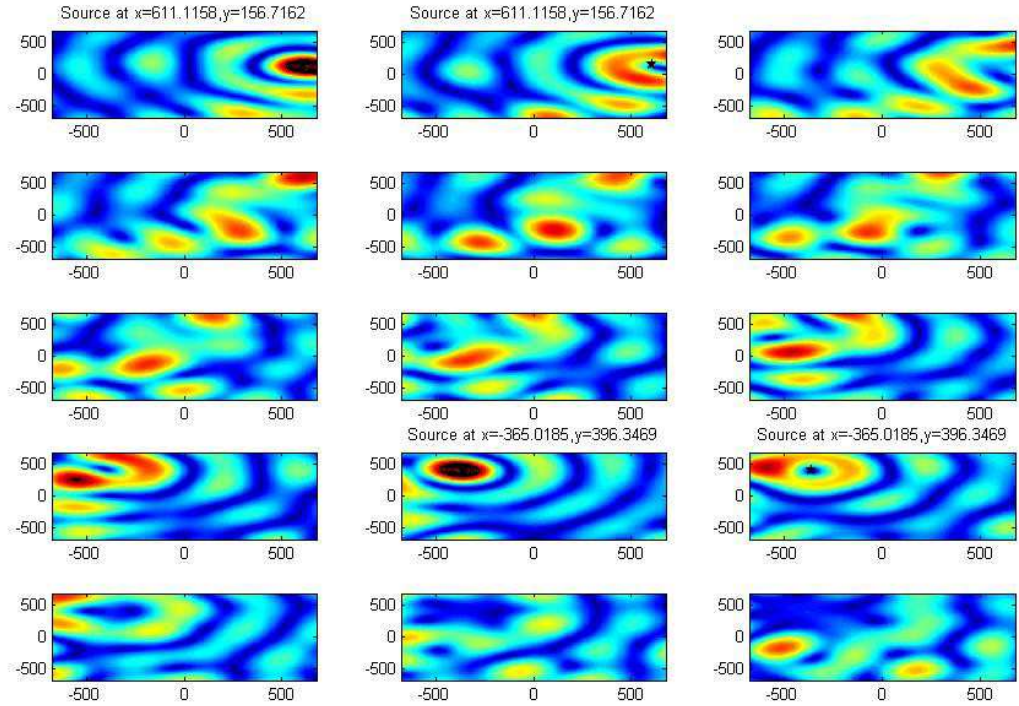}
\caption{Sky maps of $|K(x,y,k)|/\eta$ for $\eta=3\times 10^{-4}$, $\rho=15$ and sources set apart by a minimum of 10 frequency bins. One 
frequency bin has size $\delta f = 3.17\times 10^{-8}$ corresponding to  $1/T_{\rm obs} $ where $T_{\rm obs}=1$~year is the
period of observation.
 Contours are drawn at the levels $[4,4.2,4.4,4.6,4.8,5.0]$. The frequency $k_j$ corresponding to the panel in row $l$ and column $m$, from
top and left respectively, is given by $k_{1,1}+((l-1)\times 4+m)\times \delta f$.
The location of a source is indicated in the title of the panels that 
are closest in frequency on either side of the source frequency. The  '$\star$' symbol marks the source location.
\label{snr15_10_to_15_1}}
\end{figure*}
\begin{figure*}
\includegraphics[scale=1.0]{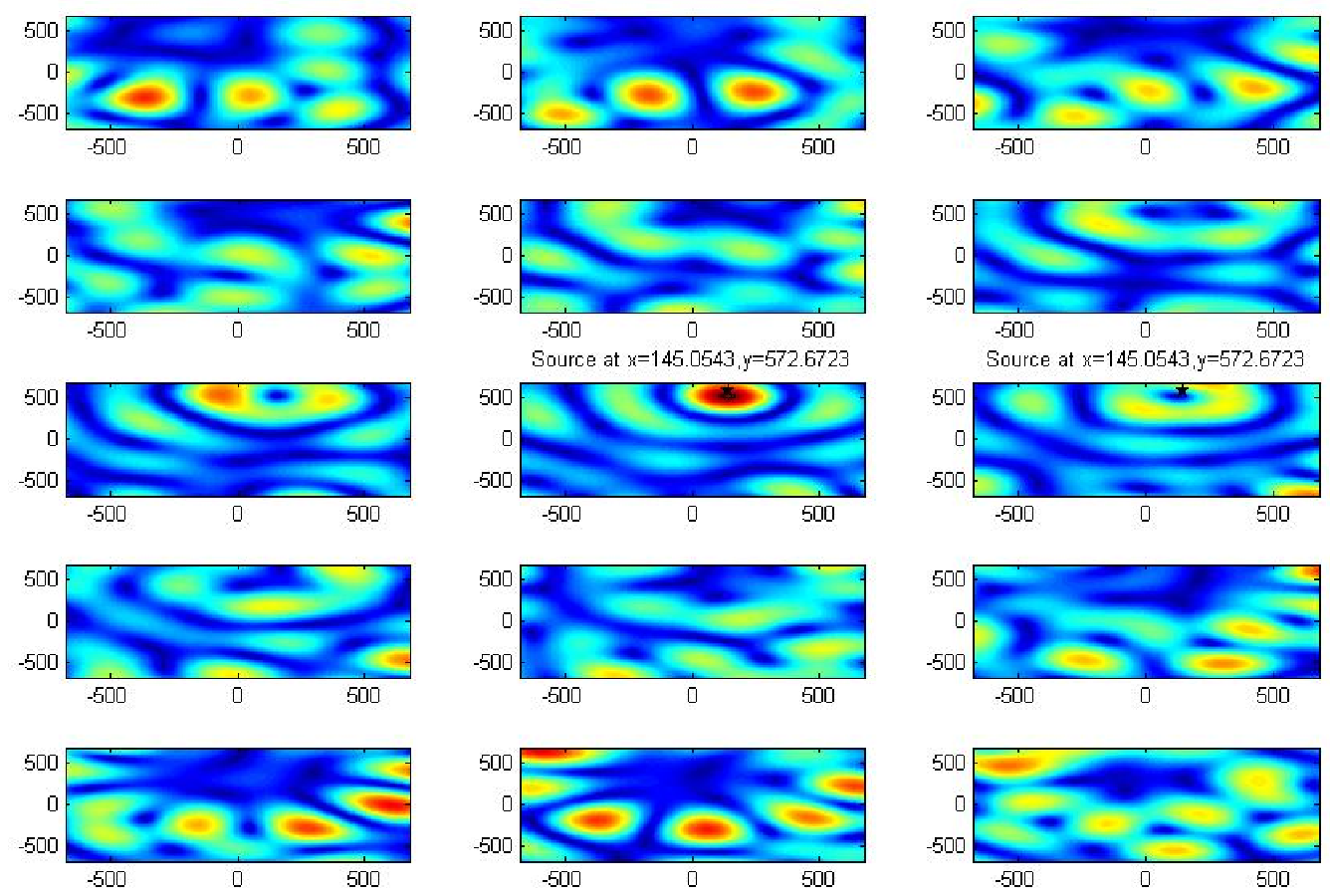}
\caption{Continuation of Fig.~\ref{snr15_10_to_15_1}\label{snr15_10_to_15_2}}
\end{figure*}

\paragraph{Noise plus signal from well isolated, weak binaries --}
Fig.~\ref{snr7_10_to_15_1} and~\ref{snr7_10_to_15_2}
 show $|K(x,y,k)|/\eta$ for a realization of LISA response that has a binary population signal $h(t)$ added to 
 the same noise realization as Fig.~\ref{noise_reconstruction_singlepanel}. For the signal we choose $\rho=7$, $b_{\rm min}=10$
and $b_{\rm max}=15$. 

After normalization with $\eta$, a threshold of $|K(x,y,k)|/\eta =1.8$ is applied to all the sky maps. The contours
of $|K(x,y,k)|/\eta$ greater than this threshold are superimposed on the sky maps. We find that, for this threshold,
 the maxima
associated with the true source locations cross the threshold. A few false peaks emerge due to the noise.
This example provides an estimate of the minimum signal strength required for a probability of detection near unity at a rate
of false source detection that is comparable to the rate of occurrence of true binaries. 
\begin{figure*}
\includegraphics[scale=1.0]{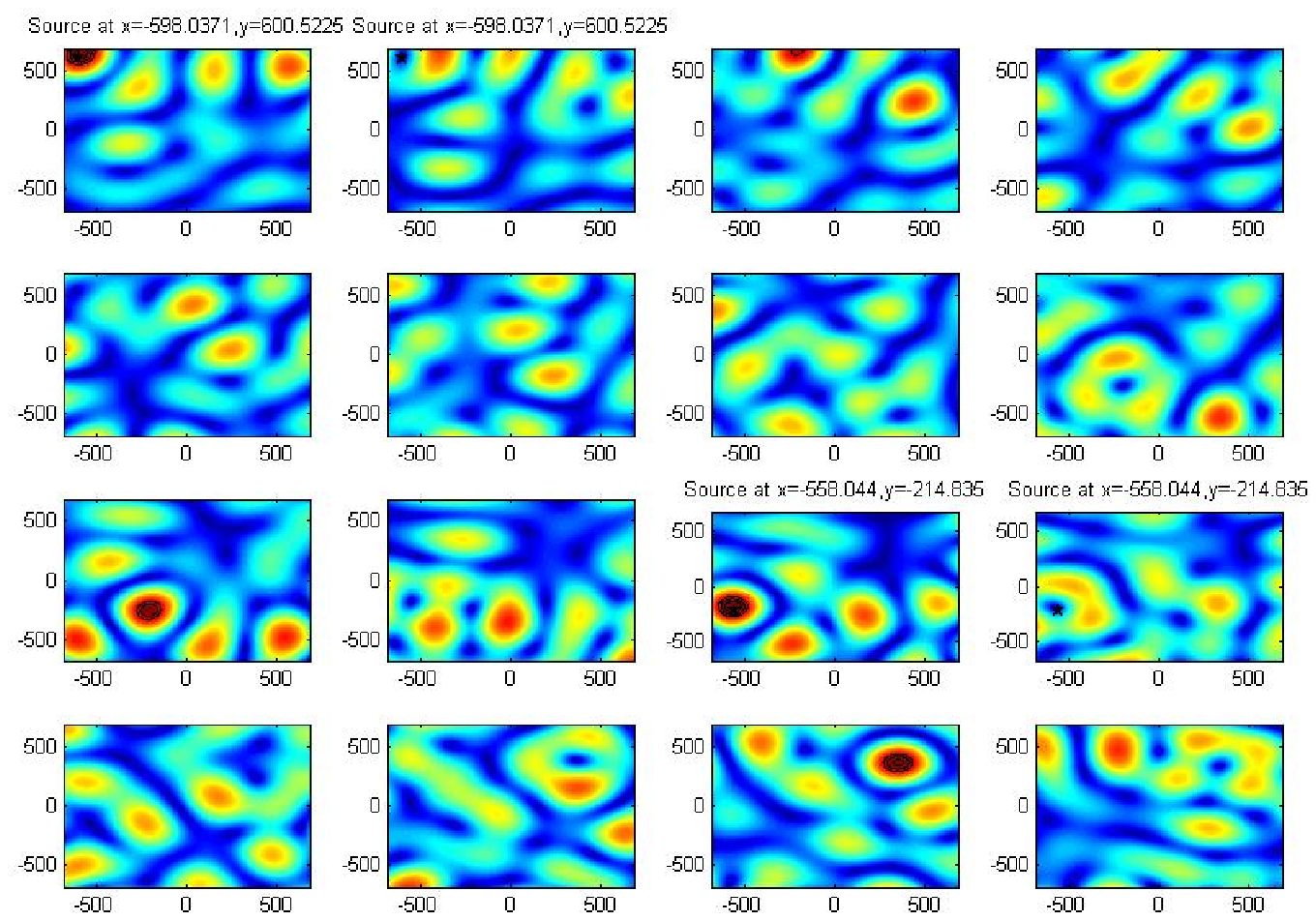}
\caption{Sky maps of $|K(x,y,k)|/\eta$ for $\eta=3\times 10^{-4}$, $\rho=7$ and sources set apart by a minimum of 10 frequency bins. One 
frequency bin has size $\delta f = 3.17\times 10^{-8}$ corresponding to  $1/T_{\rm obs} $ where $T_{\rm obs}=1$~year is the
period of observation.
 Contours are drawn at the levels $[1.8,1.9,2.0]$. The frequency $k_j$ corresponding to the panel in row $l$ and column $m$, from
top and left respectively, is given by $k_{1,1}+((l-1)\times 4+m)\times \delta f$.
The location of a source is indicated in the title of the panels that 
are closest in frequency on either side of the source frequency. The  '$\star$' symbol marks the source location.
\label{snr7_10_to_15_1}}
\end{figure*}
\begin{figure*}
\includegraphics[scale=1.0]{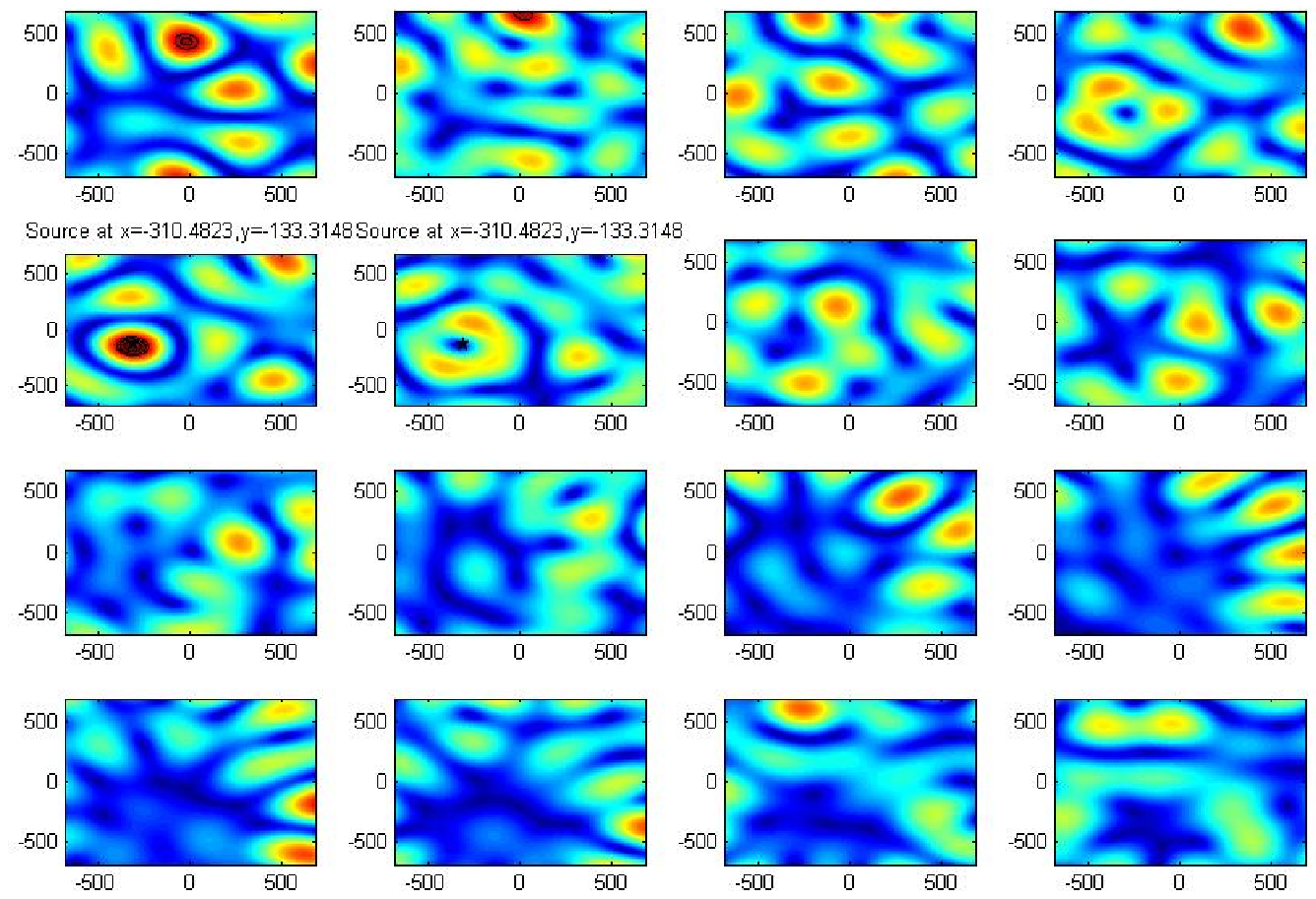}
\caption{Continuation of Fig.~\ref{snr7_10_to_15_1} in frequency. \label{snr7_10_to_15_2}}
\end{figure*}

\paragraph{Noise plus signal from close, weak binaries --}
Fig.~\ref{snr7_4_to_8_1} and~\ref{snr7_4_to_8_2}
 show $|K(x,y,k)|/\eta$ for a realization of LISA response that has a binary population signal $h(t)$ added to 
 the same noise realization as Fig.~\ref{noise_reconstruction_singlepanel}. For the signal we choose $\rho=7$, $b_{\rm min}=4$
and $b_{\rm max}=8$. After normalization with $\eta$, a threshold of $|K(x,y,k)|/\eta =1.8$ is applied to all the sky maps. The contours
of $|K(x,y,k)|/\eta$ greater than this threshold are superimposed on the sky maps. 

In this example we see the strong effect of the sidelobes of the psf. The number of false sources increases compared to
the earlier example. However,  the 
true source locations continue to have strong maxima above the threshold. A deconvolution method should be used 
in this situation to mitigate the effect of the sidelobes.
\begin{figure*}
\includegraphics[scale=1.0]{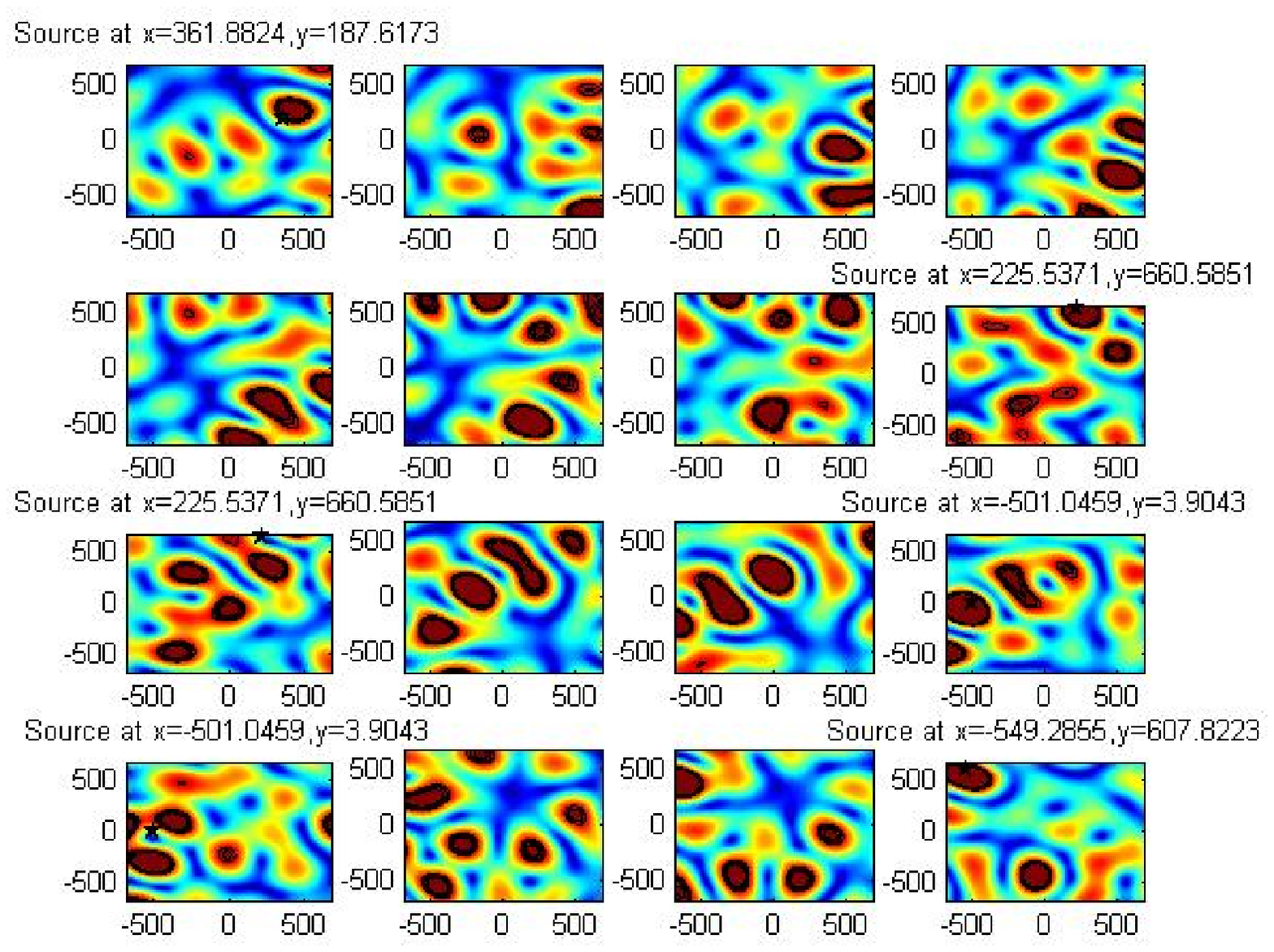}
\caption{Sky maps of $|K(x,y,k)|/\eta$ for $\eta=3\times 10^{-4}$, $\rho=7$ and sources set apart by a minimum of 4 frequency bins. One 
frequency bin has size $\delta f = 3.17\times 10^{-8}$ corresponding to  $1/T_{\rm obs} $ where $T_{\rm obs}=1$~year is the
period of observation.
 Contours are drawn at the levels $[1.8,1.9,2.0]$. The frequency $k_j$ corresponding to the panel in row $l$ and column $m$, from
top and left respectively, is given by $k_{1,1}+((l-1)\times 4+m)\times \delta f$. The location of a source is indicated in the title of the panels that 
are closest in frequency on either side of the source frequency. The  '$\star$' symbol marks the source location.
\label{snr7_4_to_8_1}}
\end{figure*}
\begin{figure*}
\includegraphics[scale=1.0]{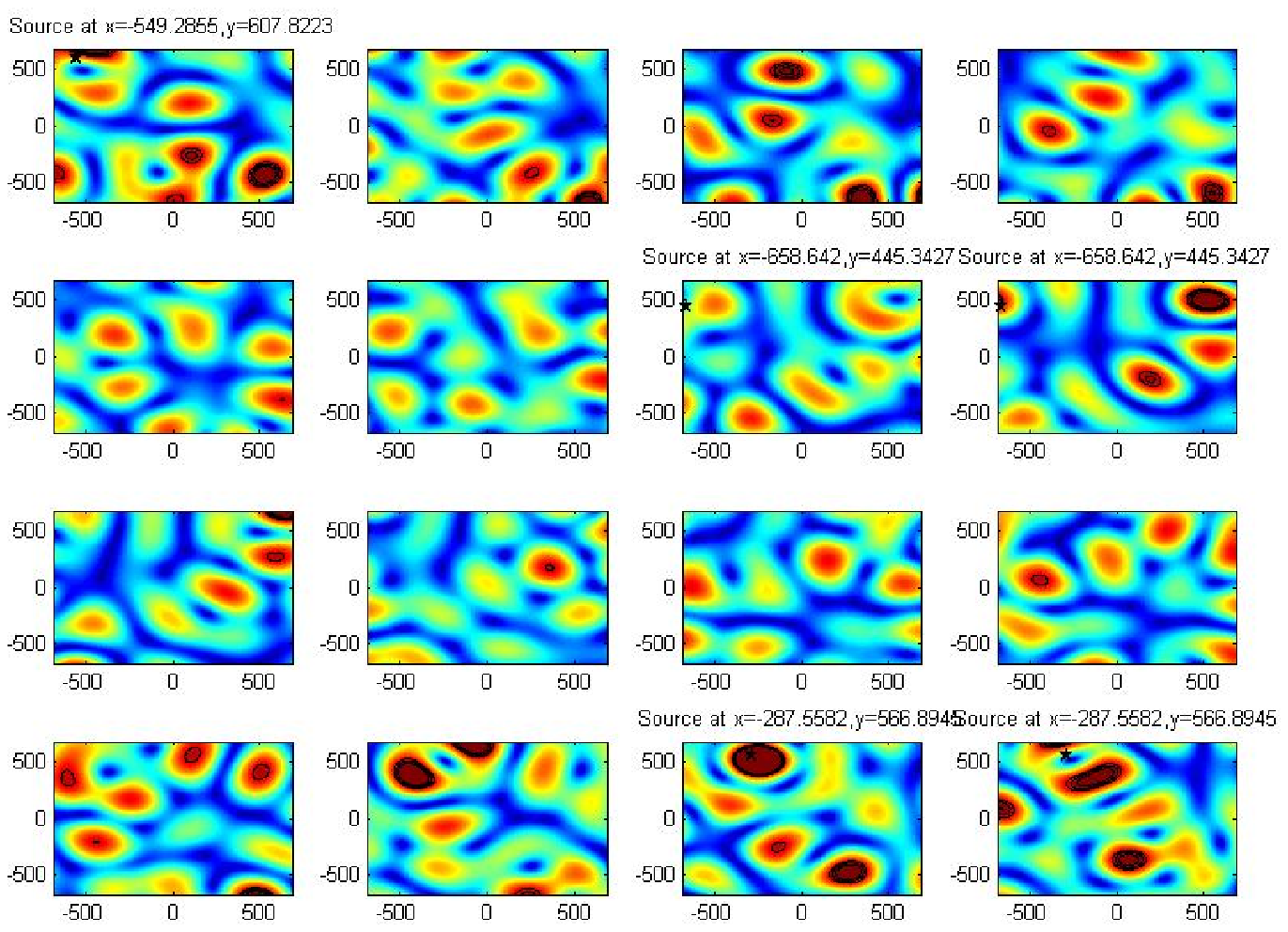}
\caption{Continuation of Fig.~\ref{snr7_4_to_8_1}\label{snr7_4_to_8_2}}
\end{figure*}

\section{Conclusions}
 We present the first
steps of a tomographic approach to the reconstruction of the LISA Galactic binary distribution. The connection between the LISA response and 
the Radon transform of the binary distribution function is established. The direct Fourier inversion method is then applied to
invert the response and reconstruct the distribution function. 

A full characterization of the performance of the method will require Monte Carlo simulations with realistic LISA
data. In this paper, we have presented
exploratory results, obtained using a simple source identification procedure.  We find that near $2$~mHz, it is possible
to recover binary sources having an snr of  about 7 and separated by about 10 (one year) frequency bins. This is close to
the expected density of binary sources around this frequency. For a higher density of sources,
  one must contend with a large number of false sources that arise from the superposition of the sidelobes of 
the psf. Deconvolution methods such as CLEAN should be useful in
 reducing this effect.

The reconstruction method presented in the paper does not take into account (i) the amplitude and phase modulations due to the 
rotating antenna pattern of LISA, and (ii) the presence of two interferometers in the LISA configuration. Inclusion of these 
factors into the method is straightforward and doing so should lead to better performance. 

Since the nature of the orbit of LISA
does not furnish a sufficient number of projections for exact inversion, we had to take recourse to a simple approximation in order
to apply the Direct Fourier inversion method.  This  situation of missing projections is 
 encountered quite often in real world imaging
applications, though probably not at the severe level of LISA.   Thus, a better approach to inversion may be possible than
the straightforward one used here. Work on this and the issues mentioned above is in progress.

The attractive features of the tomographic approach are the low and fixed computational cost of the Radon inverse transform and
the linearity of the method. For 1~year of data sampled at 0.1~Hz, a typical Pentium-4 desktop needs about 2 hours to 
produce the reconstructions shown in this paper. This is independent of the number of sources but does depend on the 
number of frequencies at which a sky map is obtained. However, the method can be parallelized quite easily and, hence, computational
cost does not appear to be a significant issue.

\section*{Acknowledgements}
This work was supported by 
NASA grant NAG5-13396 to the Center for Gravitational Wave Astronomy at the University of Texas at Brownsville.
We thank  L.~S.~Finn and M.~Benacquista for wide ranging discussions regarding
 the LISA Galactic binary problem. SDM would like to acknowledge discussions with C.~Cutler and T.~Prince.

\begin{appendix}
\end{appendix}


\begin{thebibliography}{1}
\bibitem{LISA}
P.~Bender {\em et al},``LISA: A Cornerstone Mission for the Observation of Gravitational Waves",
System and  Technology Study Report ESA-SCI(2000) 11,  2000.
\bibitem{BH97}
Bender P. L. and   Hils D.,{\it Class. Quant. Grav.}, {\bf 14}, 1439(1997).
\bibitem{CL2003}
N. J. Cornish  and S. L. Larson,
{\it Phys. Rev.},  {\bf D 67}, 103001 (2003).
\bibitem{CR2005} N.~J.~Cornish, J.~Crowder, {\em Phys.~Rev.} D {\bf 72} 043005 (2005).
\bibitem{RU2005} R. Umst\"aatter	{\it et. al.} , {\it Phys. Rev.} D {\bf 72}, 022001 (2005).
\bibitem{RD1917}Radon J, {\it Ber. Verh. S\"achs. Akad. Wiss. Leipzig, Math. Phys. K1}. {\bf 69}, 262 (1917).
%\bibitem{TDI}
%	J.W.~Armstrong, F.~B.~Eastabrook and M. Tinto, \/{\it Ap.\ J.}\, {\bf 527}, 814 (1999);\\
%	M Tinto and S V Dhurandhar, gr-qc/0409034.
\bibitem{hoegbom} J.~H\"ogbom, {\em Astrophys. J.~Suppl.~Ser.} {\bf 15}, 417-426 (1974).
\bibitem{BW1956}
	Bracewell R N, {\it Aust. J. Phys}, {\bf 9}, 198(1956).
\bibitem{steeghs} D.~Steeghs, {\em Astronomische Nachricten} {\bf 325}, p.185-188 (2004); arXiv:astro-ph/0312170.
\bibitem{RNTRNS}
	Deans S R, {\it `` The Radon Transform and Some of its Applications''}, 
	John Wiley \& Sons(1983).
\bibitem{BR1967}
	R. N. Bracewell, A. C. Riddle, {\it Astrophys. J}. {\bf 150},  427(1967)
\bibitem{RAMLM71}
	G. N. Ramachandran, A. V. Lakshminarayanan, {\it Proc. Natl. Acad. Sci}. 
	{\bf 68}, 2236(1971).
\bibitem{CUT98}C. Cutler, \emph{Phys. Rev.} D \textbf{57}, 7089 (1998).
\bibitem{AS72} M.~Abramowitz, I.~A.~Stegun, {\em Handbook of Mathematical Functions} (Dover: New York, 1972).
\end{thebibliography}
\end{document}